# Student and Faculty Adviser Insights in an Agile Methodology Integrated Filipino Company-Sponsored I.T. Capstone Program

Giuseppe Ng[1] and Rey Vincenzo Cruz[2]

[1]Department of Information Science and Technology, University of Asia and the Pacific, Pasig, Philippines
[2]Department of Mathematics, University of Asia and the Pacific, Pasig, Philippines
[1]giuseppe.ng@uap.asia, [2]vincenzo.cruz@uap.asia

*Abstract*

*To improve the Information Technology (I.T.) graduate skill set, students need to be immersed in as realistic a software development environment as possible. In continuing our work on integrating Agile Methodology into the Capstone Program of our Bachelor of Science in I.T. (BSIT) degree program, this paper discusses the student challenges and difficulties during the software development project, and provides recommendations on improving the student overall learning process in such a program. We collected survey data from the whole population of 90 BSITstudents across four academic years about their experience with their client and the Capstone Program itself. Conceptual content analysis was then applied to discover and describe underlying themes. Also, faculty advisers were tasked with writing about their interactions, thoughts, and observations on their respective student group advisees. These showed issues with time management, communication, and competency. Also, groups that excelled exhibited better team coordination and a complete grasp of the Agile methodology. For future implementations, clearer task definition and reducing the skill gaps are necessary for better execution.*

## 1. Introduction

Previous works have established that Information Technology (I.T.) graduates lack sufficient practical skills for I.T. industry work [1][2][3]. To help address this problem of the lack of skills, the Bachelor of Science in Information Technology (BSIT) program required its students to undergo a Capstone Program designed to have project-based approaches with direct hands-on experiences so that students can be exposed to real issues and can pick up valuable industry experience [1][4]. Agile Methodologies, which is a widely used approach to develop software in the industry, has been integrated in the teaching of software engineering to replicate realistic work environments as close as possible [4][5]. Selecting Agile was deemed more effective and more efficient than the Waterfall process in the academic setting primarily because Agile allows students to learn how to deal with changing requirements [1],

---

[6][7]. In addition, students can pick up soft skills for effective communication with real clients, learn how to improve teamwork among members, and problem solving and critical thinking skills - skills consistently included among the major 21st century skills frameworks [8][9][10].

From previous studies, client dissatisfaction with student output had been a challenge, leading to skepticism and lack of commitment from clients in the Capstone Program [11][12][13][14]. However, in the University of Asia and the Pacific (UA&P), our ongoing study into integrating Agile Methodologies into the BSIT Capstone project course has already shown positive effects on product output from the point of view of the clients [15]. Our previous paper [15] showed team communication and coding competency as factors to consider; nevertheless, it is our desire to look deeper into this area, not just by identifying the gaps and challenges that students encounter, but also to make recommendations to properly prepare future students.

Software engineering is defined as a structured way of creating software solutions within time and resources. Agile Methodologies is seen as a way of building and releasing software through an incremental and iterative form of development and release [16]. This is achieved through close client collaboration and focusing on the software product, as emphasized by the Agile Manifesto [16][17][18].

The design of the UA&P BSIT Capstone program has already been covered in-depth in our previous works [15]. Instead, we would like to focus our paper on the previous recommendations and the challenges that remain. The following are the research questions (RQ) that we are focusing on:

**RQ1.** What were the challenges students identified with working on their respective capstone projects?

**RQ2.** What were the group-related difficulties students faced during capstone?

**RQ3.** What were effective groups doing that others were not able to?

**RQ4.** What recommendations can be made to improve consistency of student performance in capstone?

## 2. Related work

There have been various studies of Agile Methods in capstone programs in the past focusing on the project execution performance. The work by Fan [9] defined roles in such an integrated program, capstone coordinator, Capstone instructor, Faculty advisor, and Industry mentor, to provide proper support to student groups. In the study, they highlighted the challenge of industry mentors' unavailability. Due to such circumstances, in addition to employing a digital collaborative platform, a tweak to the release evaluation cycle was done. In the event that the industry mentor was not available, the students could get students from other groups to provide additional insights on the progress.

Fagerholm and Vihavainen [8] proposed an evaluation framework focusing on several factors: presence, activity, eagerness, devotion, contribution, expert maturity, process and result. This was captured through the use of surveys. Using such evaluation metrics can help assess the group's teamwork.

The capstone program in [4] combined Scrum and Kanban concepts. Incorporating Scrum and Kanban boards gave students flexibility in how the projects were executed. Their research also proposed the integration of lectures to supplement students' gaps in employing Scrum and Kanban.

The results in [6] pointed to various impacts of utilizing digital platforms in their Agile Methods integrated course. They noted that the use of virtual meetings instead of personal meetings caused negligence on some group members. Students cited the project management challenges but that scrum encouraged more engagement from other student group members. The frequent release of software meant that students had a better understanding of the software they were building.

The work in [1] put emphasis on the student learnings from such a program. Through the use of surveys throughout the duration of the various projects, students were found to have underestimated the teamwork-related challenges. In addition, technical challenges and communication challenges that had been evident at the start of the project also diminished over time. The result of the study highlighted the relevance of soft skills.

As with our previous work [15], we maintain a post-positivism worldview where the knowledge base of the researcher may influence the results of the research through existing knowledge of Agile methodology and its underlying theories [19]. We also maintain a descriptive research approach as external observers of our students' interaction with realistic industry environments [20].

## 3. Methodology

In order to preserve consistency in the data being collected, we maintain the same evaluation framework in our previous work [15] which involved a conceptual content analysis with mixed data collected via surveys, interviews and observations. The students maintain constant communication with their clients with a formal survey evaluation at the end of each sprint. This is consistent with the works from [8]. Aside from this project-focused performance measurement tool, we also employ the other methods to capture student experiences.

Online collaborative platforms such as Slack, and Trello are employed to help students and faculty advisers manage the software project and its progress. This is consistent with the works from [6] and [9].

Assigned faculty advisers are tasked with mentoring the student groups and also observing their performance. When deemed necessary, faculty advisers communicate with students in need of help directly to better understand their circumstances and provide guidance. Scrums are also opportunities for the group to manage the project as well as consult the adviser. The use of scrum is in reference to [4]. The end of sprint serves as a formal checkpoint where faculty advisers can assess students, tracking their progress and experiences through observation. The use of Adviser roles mirrors the works from [9].

Faculty adviser insights are recorded in a journal for post mortem analysis. In addition, a post program survey is sent out to students for feedback on their challenges. The use of surveys as evaluation tools is consistent with [8].

We apply conceptual content analysis on the survey to students and revisit previous results to draw meaningful themes. [Table 1] illustrates the Data analysis plan.

Table 1. Data analysis plan

| Respondents | Data Gathering Instruments | Data Analysis |
|---|---|---|
| Students | Student survey | Conceptual Content Analysis |
| Faculty Advisers | Observational notes | |

The student feedback and faculty adviser observations were encoded into a spreadsheet. Each of the comments were read, categorized and re-categorized until a status quo was reached. An independent observer reviewed the work to ensure the validity of the categories.

## 4. Research setup

In total, we have gathered data from all 90 BSIT students across 4 senior I.T. batches from Academic Years 2016 to 2020. The breakdown of students is shown in [Table 2].

Table 2. Student population and adviser assignments

| Academic Year | No. of Students | No. of Groups | No. of Advisers |
| --- | --- | --- | --- |
| 2016-2017 | 31 | 10 | 1 |
| 2017-2018 | 18 | 6 | 1 |
| 2018-2019 | 30 | 10 | 3 |
| 2019-2020 | 11 | 4 | 1 |

Student survey forms have been collected for each of the batches. Each group had been assigned a faculty adviser whose notes and observations have been collected. Adviser assignments have been done considering faculty loading and availability with respect to the student population. The respondents of this study are students, faculty advisers, and client representatives. On the capstone briefings, it was made clear that the client representatives would be evaluating the students' work. It was also made clear to the clients that their representatives may reach out and inquire with the capstone coordinator and the faculty advisers should they have any issues with the evaluation tool and the program.

The surveys utilized for students and clients are different. The student survey had three questions that have students enumerate the challenges and difficulties they faced individually and as a group. Two questions had students also list down the skills they learned and also to do an approximation of the task complexity. In addition, they were asked to describe their interactions with the client representatives.

The client survey focused on evaluating the group and the system built. The evaluation questions are structured to explain the criteria of the system. They were also requested to let us know should there be any issues with the evaluation. Details about the evaluation form are discussed in our previous work [15].

Utilizing the student forms and conceptual content analysis, we could identify the common themes across the student responses. These were then confirmed to be consistent with the observations of the faculty advisers and capstone coordinator. These could be used to answer RQ1, RQ2, and RQ3. With the extracted themes, recommendations could be drawn to answer RQ4.

As the capstone program is designed with Agile methodology in mind, student groups adopted this framework as a requirement of the program.

## 5. Research findings

In this section, we report on our findings and answer the research questions. Comments by students have been edited for the sake of brevity.

**RQ1.** What were the challenges students identified with working on their respective capstone projects?

The first key issue was regarding the changes of their initial design compared to the final version of their software. Applying conceptual content analysis on the student responses showed interesting themes that are listed in [Table 3].

Table 3. Additional changes on capstone initial designs

| Theme | Student Response |
|---|---|
| Lack of understanding of the business process | Many functionalities were modified due to better understanding. |
| | It was hard to grasp the initial system leading to changes. |
| | Our impression of the software was so far from the actual software leading to big changes |
| Lack of experience in building systems | I did not expect the system to be difficult. I think we underestimated it. |
| | The initial design was just a big picture of what the system should be made to do. |
| | It was too simple compared to the software delivered. |
| Client requested new features and changes | We realized that the client didn't really know what they wanted. But prototypes demonstrations helped them to suggest new features. |
| | There were many changes since clients requested for more features |

As shown in the table, three basic themes were drawn. First is a lack of understanding of the business process itself. When considering that students were expected to make a system to address the client's concerns, the fundamental understanding of the basic operations and the client's business process was crucial.

Given that the students only had a fixed time window to create their initial design, there were gaps in their understanding. This led to missing functionalities, if not incorrect or inadequate design. Such issues led to additional system corrections and modifications.

Second, as students did not have as much experience in building major systems, the responses showed that they have underestimated the task. The lack of experience also led to overly simplistic system designs. This meant that gaps in the system functionalities led to corrections and additional work.

Lastly, the nature of agile methodologies of having close client collaboration lent itself to changes and, at times, realignment of work. This has been pointed out in our previous work [15] and other references [18][21]. As students showed their progress, the clients would have additional requests and new features to improve the system.

**RQ2.** What were the group-related difficulties students faced during capstone?

In assessing the feedback from students, we could see that there was a disparity of skills among members of the groups. From our conceptual content analysis, we identified these issues: (1) Time Management Skills, (2) Coding Competency, (3) Team Communication. Table 4 illustrates our content analysis regarding time management skills.

Table 4. Time management issues

| Theme | Student Response |
|---|---|
| Balancing Academic Load | The school load was a problem because some members were overloaded in terms of units. |
| | Balancing academic workload was a challenge. |
| Extra-Curricular Activities | My challenges personally was that I took other commitments while working on capstone. |
| | Some challenges were student org related. |
| Respecting Deadlines | Too lenient with working on deliverables |
| | We liked to cram our deadlines. Inconsistent work output. There were moments when nobody worked on the project. |
| | Deadlines are extended well beyond acceptable times early in the project. |

As can be seen, academic load was a significant factor when managing capstone. It was an even bigger challenge for students who were taking more units than usual. This was mainly due to students being delayed curriculum-wise. In one case, a student's class schedule extended until evenings which limited the time to work on the project.

From the observation of faculty advisers, it was worth noting that the midterm exams typically lined up with project's sprint two. Furthermore, final exams and other semester long project deadlines usually lined up with the last sprint. For a few students, they were also managing internship responsibilities along with capstone.

Extra-curricular activities was another factor to consider. Students still engaged with other student organization activities, and some of them prioritized these activities over the project tasks. Hence, students extended their task deadlines after missing them. Cramming was a subsequent reaction to closing deadlines, often at the expense of product quality, as observed by faculty advisers.

In terms of competency issues, student responses gave us more insight over factors that contributed towards coding weaknesses. This is shown in [Table 5].

Table 5. Coding competency issues

| Theme | Student Response |
|---|---|
| Some of the members were delayed curriculum-wise | Students need coding skills before entering the capstone program. |
| | One of the members had a hard time programming because he hasn't learned the underlying skills. |
| | I had to study topics I have not taken yet to work on capstone. |
| Skill set disparity caused varying workloads and affected progress. | Some tasks needed to be redone due to incorrect or not working code. |
| | Lack of programming skills is evident from some members Imbalanced coding skills among members causing major team adjustments. |
| | Coding skills is my main issue. Finding good resources and how to use them have been difficult for me. |
| Teaching and guiding other members caused delays | We had to devote time to teach members who weren't proficient. |
| | Very difficult to teach groupmates who are not skilled yet in coding. |

Given that a few of the student responses pointed to researching good resources to use for tasks, students were asked to rate from 0 to 10, what percentage of total time spent on the project ticket task was devoted to research. A rating of 0 means none at all while a rating of 10 means all of the student's time. Feedback showed that when they had time to work, a significant amount of time had been spent on researching solutions and ideas on how to address this problem. [Figure 1] shows that most of the students rated 7 and 8 for the time devoted for researching. This showed that for most students, there was still overhead of researching a task instead of directly writing the code.

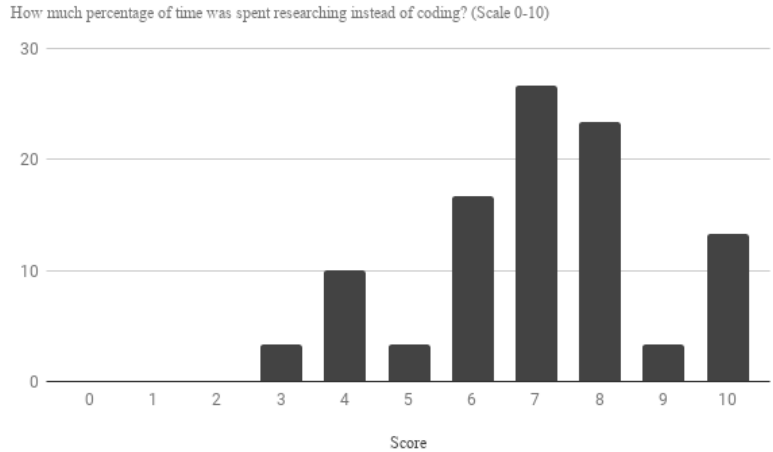

Figure 1. Percentage of time researching instead of coding

Noted as well by faculty advisers, many students focused only on the success path of a particular function instead of checking possible faults. This decreased the quality of the software and introduced more bugs that students needed to revisit later on. This characteristic is consistent with the insights in [22].

Team communication was another aspect uncovered by the student responses. From our conceptual content analysis, we identified two major themes. First was an overall lack of teamwork where teams did not check the state of other tasks. At times, members were preoccupied by other tasks.

Table 6. Team communication issues

| Theme | Student Response |
|---|---|
| Lack of Teamwork | Communicating was difficult because members were busy with other unrelated matters. |
| | We didn't have teamwork because we rarely checked on each other's progress. Team coordination. The members always have excuses when they cannot deliver tasks or when they are late. |
| Frequent unexplained absences | Sometimes group members are unavailable without notice. |
| | Usually, the group was incomplete during client meetings leading to unaddressed discussion points. |

Second was attendance. Some group members were missing during meetings. As noted by faculty advisers, some groups were frequently incomplete during scrum meetings affecting the efficacy of the meeting. Some client meetings happened without all the members being present causing incomplete information being escalated to the next sprints. These were also pointed out in our previous work [15]. Table 6 lists down the team communication issues that we identified based on the students' responses. Communication issues and technical challenges are consistent with the results from [1].

**RQ3.** What were effective teams doing that others were not able to?

In referencing the student feedback, we also studied the groups that performed well. Identification of these groups were based on the client evaluation and faculty adviser observations. For these groups, the task tickets were completed in a timely manner with every scrum showing steady progress as determined by their respective faculty adviser and the capstone coordinator. In terms of client evaluation, the groups had impressed client representatives as affirmed in their written feedback.

These groups showed: (1) proper division of work, (2) proper grasp of agile concepts, (3) communication, (4) proper team support, (5) effective use of collaboration tools. (1), (3), and (4) are all consistent with the skills pointed out in Dede's work [10]. These are listed in [Table 7].

Table 7. Effective team characteristics

| Theme | Student Response |
|---|---|
| Proper division of work | I divided my team based on our skills. Our shared strength is by working on our skill that we are good at to finish all the tasks on time. |
| | We were able to do a lot of things well based on our strengths. |
| | The capstone group has a great chemistry because we already know each other too well and because of this, we were able to distribute major tasks easily. |
| Proper grasp of agile concepts | Scrums were useful because we were able to track each member's progress and discuss issues. |
| | Scrum was great that since the progress are being monitored |
| | All of our sprints were done well. We always targeted for at least 30 story points per sprint. |
| | Agile methods helped us plan our tasks better. |
| | Scrum meetings is a good time to seek help when there are bugs or problems that we can't solve. |
| Communication | My group and I were able to communicate properly on what tasks we want to work on. |
| | We were able to meet and go to scheduled meetings together, and if someone can't go that person usually informs us. |
| | My team and I are transparent on the things that we know and don't know. |
| Proper team support | I built the back end architecture in order for my teammates to develop functionalities easily. |
| | As much as possible, we try to help each other or take that person's task to hasten the progress of our system. |
| | Willing to self-learn when needed |
| Effective use of online messaging | Slack eliminated the distraction of unrelated stuffs like videos or posts that Facebook and other social media sites have. |
| | Slack is very effective since there are no distractions unlike Facebook. |
| | Bitbucket integration notifies the team when someone pushed code. |
| Effective use of online project management | The Trello board really helps in breaking down and managing tasks. |
| | Trello did a great job in organizing our plans. |
| | Our tasks had enough breakdowns to ensure that every task is fulfilled at a given deadline. |

First, there was proper division of work as the group managed to distribute the tasks to utilize the strengths of each member. This, in turn, meant that these groups were able to support each member completely. This was shown by the ability of every member to self-learn and also assist each other. These groups exhibited agility in adjusting based on circumstances.

By showing that the groups were agile and flexible, these effective groups showed that they understood the agile concepts. They were able to use the sprint planning to plot the course of action but still remain nimble to adjust to issues that arise. These effective groups also managed to utilize scrum effectively. These insights are consistent with the results from [4] where scrum helped students have flexibility in project management.

Effective utilization of online platforms such as Slack and Trello allowed the groups to manage and adjust their project tasks. This in turn improved the communication between members. Members were able to coordinate on meetings, and task delegation.

**RQ4.** What recommendations can be done to improve consistency of student performance in capstone?

Based on our findings, there are several issues that need to be addressed. First is to increase the efficiency of the coding activity by reducing the research effort per task. We believe that this can be addressed in the sprint planning phase where members can and should be discussing implementation details of every task for the sprint. Before this, proper identification of the significant knowledge, skill, abilities and other related characteristics (KSAO) of individual team-members will have to be done. By annotating the task tickets with ideas from each member, the task will have enough details so as to effectively reduce the overhead of research work primarily by aligning team-member KSAO to efficiently resolve the identified task demands [23]. This also mirrors the agile practice that is explained in [22] and addresses vagueness of the tasks in Table 3 and the communication issues as noted in [Table 6]. Analyzing the tasks and its complexity supports the work in [6].

The second issue regards the group competency in terms of programming. We advise that groups establish operating protocols and support structures for members who need assistance. This involves establishing communication procedures when members are unable to progress on their tasks due to coding issues. We have already tested this idea informally with some of the groups with skill imbalance through establishing which communication channels to use when they are unable to proceed and to coordinate working hours amongst themselves. These address the skill and communication issues listed down in [Table 5] and [Table 6]. These should lead to determining task demand specific training interventions and putting them in place. We find this approach has potential and could be a topic for further research work.

The third issue regards time management skills of students. While we discourage faculty advisers from micromanaging every group and every member, we believe they can still provide useful suggestions to aid students in this issue by organizing workshops. This addresses the time management issues that were shown in [Table 4].

One faculty adviser suggested that a group leader, chosen by the group members or faculty-in-charge, be formally identified and duly recognized by the client and faculty adviser for when the need for executive decision arises. Per [23], this leader should also perform instructional and regulatory functions. Another suggestion was having the faculty adviser be given more insight on how to better manage scrum meetings in order to reap the most from them.

## 6. Other concerns

In this section, we take note of several special circumstances. These issues are outliers in the data collected, however, we have deemed it necessary to mention them as they significantly affected the final outputs.

The nature of the Capstone Program requires a client, but in one group, the client could no longer wait for the group's output and procured their own solution. This meant that the group of students were forced to change topics. This was a circumstance of misaligned expectations as the client was not made aware of when the students would be starting and completing the actual implementation.

In one group, health became a major concern. This was due to members becoming ill for prolonged periods of time and caused project progress to slow down. This risk was managed with the help of the faculty adviser in the weekly scrums that allowed the members to assess and plan around the limitations. This involved identifying how long to wait for the member's recovery and what work could be done in preparation of taking the tasks from the ill member.

In another group's case, a member had unexpectedly taken leave of absence, impacting the composition of the group and the plan taken. The risk was mitigated with the help of the faculty adviser during scrum. This involved identifying the project scope, and knowledge transfer from the departing member. Project details that were not known by the remaining members were re-clarified.

Another issue pertains to project failures. In our previous work [15] we succinctly discussed one instance of project failure which involved lack of communication and code competency issues. This other project failure case also showed communication issues compounded by other personal issues that made it difficult for the group to function despite the support of the faculty adviser. They also failed to see the importance of the agile concepts and scrum, resulting in an unstructured way of working on tasks. This led to tasks being marked done when the code submitted had clear usability and functional issues. These are clear violations of the agile concepts and procedures established in [18][21][22].

## 7. Conclusion

Our research explored the student point of view more thoroughly than in our previous works. We identified issues regarding group competency, communication, and time management that can impact typical student groups working on large software development projects. It should be noted, however, that it is precisely through experiencing these difficulties that these 21st Century skills of leadership, teamsmanship, communication, and self-direction are developed [24]. We also identified the characteristics that make a good student group and also noted that successful groups dutifully followed the agile method of software development and recognized its value in ensuring the successful execution of their Capstone project.

For future work, we believe that designing a Capstone Program while establishing team coordination and communication protocols to close the skill gaps in student groups and improve project management would be worth pursuing to improve the quality of I.T. graduates.

## Acknowledgements

We would like to extend our gratitude to the Department of Information Science and Technology as well as all the I.T. students who were part of this research.

...

# Authors

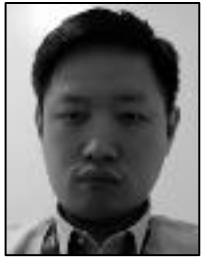

**Giuseppe Ng**

Giuseppe C. Ng is currently a faculty member of the Department of Information Science and Technology of the School of Sciences and Engineering at the University of Asia and the Pacific. He has 11 years of experience in the IT industry as a software engineer prior to teaching. He is also currently pursuing a PhD in Computer Science at De La Salle University.

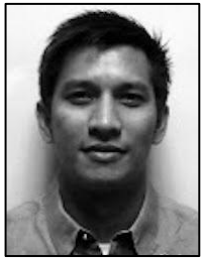

**Rey Vincenzo Cruz:**

Rey Vincenzo Y. Cruz is a faculty member of the Department of Mathematics of the School of Sciences and Engineering at the University of Asia and the Pacific. He is currently pursuing a PhD in Education major in Mathematics at the University of the Philippines Open University.